# The Epic Story of Maximum Likelihood

**Stephen M. Stigler**

*Abstract.* At a superficial level, the idea of maximum likelihood must be prehistoric: early hunters and gatherers may not have used the words "method of maximum likelihood" to describe their choice of where and how to hunt and gather, but it is hard to believe they would have been surprised if their method had been described in those terms. It seems a simple, even unassailable idea: Who would rise to argue in favor of a method of minimum likelihood, or even mediocre likelihood? And yet the mathematical history of the topic shows this "simple idea" is really anything but simple. Joseph Louis Lagrange, Daniel Bernoulli, Leonard Euler, Pierre Simon Laplace and Carl Friedrich Gauss are only some of those who explored the topic, not always in ways we would sanction today. In this article, that history is reviewed from back well before Fisher to the time of Lucien Le Cam's dissertation. In the process Fisher's unpublished 1930 characterization of conditions for the consistency and efficiency of maximum likelihood estimates is presented, and the mathematical basis of his three proofs discussed. In particular, Fisher's derivation of the information inequality is seen to be derived from his work on the analysis of variance, and his later approach via estimating functions was derived from Euler's Relation for homogeneous functions. The reaction to Fisher's work is reviewed, and some lessons drawn.

## 1. INTRODUCTION

In the 1860s a small group of young English intellectuals formed what they called the X Club. The name was taken as the mathematical symbol for the unknown, and the plan was to meet for dinner once a month and let the conversation take them where chance would have it. The group included the Darwinian biologist Thomas Henry Huxley and the so-cial philosopher-scientist Herbert Spencer. One evening about 1870 they met for dinner at the Athenaeum Club in London, and that evening included one exchange that so struck those present that it was repeated on several occasions. Francis Galton was not present at the dinner, but he heard separate accounts from three men who were, and he recorded it in his own memoirs. As Galton reported it, during a pause in the conversation Herbert Spencer said, "You would little think it, but I once wrote a tragedy." Huxley answered promptly, "I know the catastrophe." Spencer declared it was impossible, for he had never spoken about it before then. Huxley insisted. Spencer asked what it was. Huxley replied, "A beautiful theory, killed by a nasty, ugly little fact" (Galton, 1908, page 258).

*Stephen M. Stigler is the Ernest DeWitt Burton Distinguished Service Professor, Department of Statistics, University of Chicago, Chicago, Illinois 60637, USA e-mail: stigler@galton.uchicago.edu.*

Huxley's description of a scientific tragedy is singularly appropriate for one telling of the history

<div style="border:1px solid">

**Joe Hodges's Nasty, Ugly Little Fact (1951)**

$T_n = \bar{X}_n \quad$ if $|\bar{X}_n| \geq \dfrac{1}{n^{1/4}}$

$\quad = \alpha \bar{X}_n \quad$ if $|\bar{X}_n| < \dfrac{1}{n^{1/4}}$.

Then $\sqrt{n}(T_n - \theta)$ is asymptotically $N(0, 1)$ if $\theta \neq 0$, and asymptotically $N(0, \alpha^2)$ if $\theta = 0$.

$T_n$ is then "super-efficient" for $\theta = 0$ if $\alpha^2 < 1$.

</div>

Fig. 1.  *The example of a superefficient estimate due to Joseph L. Hodges, Jr. The example was presented in lectures in 1951, but was first published in Le Cam (1953). Here $\bar{X}_n$ is the sample mean of a random sample of size $n$ from a $N(\theta, 1)$ population, with $n \operatorname{Var}(\bar{X}_n) = 1$ all $n$, all $\theta$ (Bahadur, 1983; van der Vaart, 1997).*

of Maximum Likelihood. The theory of maximum likelihood is very beautiful indeed: a conceptually simple approach to an amazingly broad collection of problems. This theory provides a simple recipe that purports to lead to the optimum solution for all parametric problems and beyond, and not only promises an optimum estimate, but also a simple all-purpose assessment of its accuracy. And all this comes with no need for the specification of a priori probabilities, and no complicated derivation of distributions. Furthermore, it is capable of being automated in modern computers and extended to any number of dimensions. But as in Huxley's quip about Spencer's unpublished tragedy, some would have it that this theory has been "killed by a nasty, ugly little fact," most famously by Joseph Hodges's elegant simple example in 1951, pointing to the existence of "superefficient" estimates (estimates with smaller asymptotic variances than the maximum likelihood estimate). See Figure 1. And then, just as with fatally wounded slaves in the Roman Colosseum, or fatally wounded bulls in a Spanish bullring, the theory was killed yet again, several times over by others, by ingenious examples of inconsistent maximum likelihood estimates.

The full story of maximum likelihood is more complicated and less tragic than this simple account would have it. The history of maximum likelihood is more in the spirit of a Homeric epic, with long periods of peace punctuated by some small attacks building to major battles; a mixture of triumph and tragedy, all of this dominated by a few characters of heroic stature if not heroic temperament. For all its turbulent past, maximum likelihood has survived numerous assaults and remains a beautiful, if increasingly complicated theory. I propose to review that history, with a sketch of the conceptual problems of the early years and then a closer look at the bold claims of the 1920s and 1930s, and at the early arguments, some unpublished, that were devised to support them.

## 2. THE EARLY HISTORY OF MAXIMUM LIKELIHOOD

By the mid-1700s it seems to have become a commonplace among natural philosophers that problems of observational error were susceptible to mathematical description. There was essential agreement upon some elements of that description: errors, for want of a better assumption, were supposed equally able to be positive and negative, and large errors were expected to be less frequently encountered than small. Indeed, it was generally accepted that their frequency distribution followed a smooth symmetric curve. Even the goal of the observer was agreed upon: while the words employed varied, the observer sought the *most probable* position for the object of observation, be it a star declination or a geodetic location. But in the few serious attempts to treat this problem, the details varied in important ways. It was to prove quite difficult to arrive at a precise formulation that incorporated these elements, covered useful applications, and also permitted analysis.

There were early intelligent comments related to this problem already in the 1750s by Thomases Simpson and Bayes and by Johann Heinrich Lambert in 1760, but the first serious assault related to our topic was by Joseph Louis Lagrange in 1769 (Stigler, 1986, Chapter 2; 1999, Chapter 16; Sheynin, 1971; Hald, 1998, 2007). Lagrange postulated that observations varied about the desired mean according to a multinomial distribution, and in an analytical tour de force he showed that the probability of a set of observations was largest if the relative frequencies of the different possible values were used as the values of the probabilities. In modern terminology, he found that the maximum likelihood estimates of the multinomial probabilities are the sample relative frequencies. He concluded that the most probable value for the *desired* mean was then the mean value found from these probabilities, which is the arithmetic mean of the observations. It was only then, and contrary to modern practice, that Lagrange introduced the hypothesis that the multinomial probabilities followed a symmetric curve, and so he was left with only the problem of finding the probability distribution of the arithmetic mean when the error probabilities follow a curve. This he solved for several examples by introducing and using "Laplace Transforms." By introducing restrictions in the form of the curve only *after* deriving the estimates of probabilities, Lagrange's analysis had the curious consequence of always arriving at method of moment estimates, even though starting with maximum likelihood! (Lagrange, 1776; Stigler, 1999, Chapter 14; Hald, 1998, page 48.)

At about the same time, Daniel Bernoulli considered the problem in two successively very different ways. First, in 1769 he tried using the hypothesized curve as a *weight function*, in order to weight, then iteratively reweight and average the observations. This was very much like some modern robust M-estimates. Second, in 1778 (possibly after he had seen a 1774 memoir of Laplace's with a Bayesian analytical formulation), Bernoulli changed his view dramatically and used the same curve as a *density* for single observations. He multiplied these densities together, and he sought as the true value for the observed quantity, that value that made the product a maximum (Bernoulli, 1769, 1778; Stigler, 1999, Chapter 14; Laplace, 1774).

These and the other attempts of that time were primarily theoretical explorations, and did not attract many practical applications or further development. And while they all used phrases that could easily be translated into modern English as "Maximum Likelihood," and in some cases even be defended as maximum likelihood, in no case was there a reasoned defense for them or their performance. The most that was to be found was the superficial invocation that the value derived was "most probable" because it made the only probability in sight (the probability of the observed data) as large as possible.

The philosophically most cogent of these early treatments was that of Gauss, in his first publication on least squares in 1809 (Gauss, 1809). Gauss, like Daniel Bernoulli in 1778, adopted Laplace's analytical formulation, but unlike Bernoulli, Gauss explicitly invoked Laplace's Bayesian perspective using a uniform prior distribution for the unknowns. Where Laplace had then sought (and found) the posterior median (which minimized the posterior expected error), Gauss chose the posterior mode. In accord with modern maximum likelihood with normally distributed errors, this led Gauss to the method of least squares. The simplicity and tractability of the analysis made this approach very popular over the nineteenth century. By the end of that century this was sometimes known as the Gaussian method, and the approach became the staple of many textbooks, often without the explicit invocation of a uniform prior that Gauss had seen as needed to justify the procedure.

## 3. KARL PEARSON AND L. N. G. FILON

Over the 19th century, the theory of estimation generally remained around the level Laplace and Gauss left it, albeit with frequent retreats to lower levels. With regard to maximum likelihood, the most important event after Gauss's publication of 1809 occurred only on the eve of a new century, with a long memoir by Karl Pearson and Louis Napoleon George Filon, published in the *Transactions* of the Royal Society of London in 1898 (Pearson and Filon, 1898). The memoir has a place in history, more for what in the end it seemed to suggest, rather than for what it accomplished. The two authors considered a very general setting for the estimation problem—a set of multivariate observations with a distribution depending upon a potentially large array of constants to be determined. They did not refer to the constants as parameters, but it would be hard for a modern reader to view them in any other light, even though a close reading of the memoir shows that it lacked the parametric view Fisher was to introduce more than 20 years later (Stigler, 2007).

The main result of Pearson and Filon (expressed in modern terminology) came from taking a likelihood ratio (a ratio of the frequency distribution of the observed data and the frequency distribution evaluated for the same data, but with the constants slightly perturbed), expanding its logarithm in a multivariate Taylor's expansion, then approximating the coefficients by their expected values and claiming that the resulting expression gave the frequency distribution of the errors made in estimating the constants. They erred in taking the limit of the coefficients, in effect using a procedure that did not at all depend upon the method of estimation used and would at most be valid for maximum likelihood estimates, a fact they failed to recognize. Their last step employed an implicit Bayesian step in the manner of Gauss. When cubic and higher order terms were neglected, their formula would give a multivariate normal posterior distribution (extending results of Laplace a century earlier), although Pearson and Filon cautioned against doing this with skewed frequency distributions. A modern reader would recognize their resulting distribution as the normal distribution sometimes used to approximate the distribution of maximum likelihood estimates, but Pearson and Filon made no such restriction in the choice of estimate and applied it heedlessly to all manner of estimates, particularly to method of moments estimates.

The result may in hindsight be seen to be a mess, not even applying to the examples presented, and the approach was soon to be abandoned by Pearson himself. But it led to some correct results for the bivariate normal correlation coefficient, and it was bold and surely highly suggestive to a reader like Ronald Fisher, to whom I now turn. I have recently published a detailed study (Stigler, 2005) of how Fisher was led to write his 1922 watershed work on "The Mathematical Foundations of Theoretical Statistics," so I will only briefly review the main points leading to that memoir.

## 4. R. A. FISHER

At Cambridge Fisher had studied the theory of errors and even published in 1912 a short piece commending the virtues of the Gaussian approach to estimation, particularly of the standard deviation of a normally distributed sample. He had been so taken by the invariance of the estimates so derived, how (for example) the estimate of the square of a frequency constant was the square of the estimate of the constant, that he termed the criterion "absolute" (Fisher, 1912). But his approach at that time was superficial in most respects, tacitly endorsing the naïve Bayesian approach Gauss had used, without noticing the lurking inconsistency in even the example he considered, in that the estimate of the squared standard deviation based upon the distribution of the data, namely $\frac{1}{n}\sum(x_i - \bar{x})^2$, did not agree with that found applying the same principle to distribution of $\frac{1}{n}\sum(x_i - \bar{x})^2$ alone.

Four years later, Fisher sent to Pearson for possible publication a short, equally superficial critique of a *Biometrika* article by Kirstine Smith advocating the minimum chi-square approach to estimation (Smith, 1916). Pearson's thoughtful rejection letter to Fisher focused on the lack of a clear and convincing rationale for the method of choosing constants to maximize the frequency function, and Pearson even stated that he now thought the Pearson–Filon paper was remiss on the same count. He called particular attention to a perceptive footnote in Smith's paper that argued the case against the Gaussian method: the probability being maximized was not a probability but rather a probability density, an infinitesimal probability, and of what force was such meager evidence in defense of a choice? At least the minimum chi-square method optimized with respect to an actual metric. Two more years passed, and in 1918 Fisher discovered sufficiency in the context of estimating the normal standard deviation (Fisher, 1920); he recalled Pearson's challenge to produce a rationale for the method, and he was off to the races, quickly setting to work on the monumental paper on the theory of statistics that he read to the Royal Society in November 1921 and published in 1922.

## 5. FISHER'S FIRST PROOF

By my reconstruction, Fisher's discovery of sufficiency was quickly followed by the development of a short argument that he gave in that great 1922 paper; indeed it was the first mathematical argument in the paper. The essence of the argument in modern notation is the following. Suppose you have two candidates as estimates for a parameter $\theta$, denoted by $S$ and $T$. Suppose that $T$ is a sufficient statistic for $\theta$. Since generally both $S$ and $T$ are *approximately* normal with large samples, let us (anticipating a species of argument Wald was to develop rigorously in 1943) follow Fisher in considering that $S$ and $T$ actually have a bivariate normal distribution, both with expectation $= \theta$, and with standard deviations $\sigma_S$ and $\sigma_T$ and correlation $\rho$. Then the standard facts of the bivariate normal distribution tell us that $E(S|T=t) = \theta + \rho(\sigma_S/\sigma_T)(t-\theta)$. Since $T$ is sufficient, this cannot depend upon $\theta$, which is only possible if $\rho(\sigma_S/\sigma_T) = 1$, or if $\sigma_T = \rho\sigma_S \leq \sigma_S$. Thus $T$ cannot have a larger mean squared error than any other such estimate $S$, and so must be optimum according to a clear metric criterion, expected squared error! In one stroke Fisher had (if one accepts the substitution of exact for approximate normality) the simple and powerful result:

> Sufficiency implies optimality, at least when combined with consistency and asymptotic normality.

The question was, how general is this result? Neither Fisher nor much of posterity thought of consistency and asymptotic normality as major restrictions. After all, who would use an inconsistent estimate, and while there are noted exceptions, is not asymptotic normality the general rule? Indeed, Fisher clearly knew the result was stronger than this, that a sufficient estimate captured all the information in the data in even stronger senses; the argument was only to present the claim in terms of a specific criterion, minimum standard error. But what about sufficiency?

At this point Fisher appears to have made an interesting and highly productive mistake. He quickly explored a number of other parametric examples and came to the conclusion that maximizing the likelihood *always* led to an estimate that was a function of a sufficient statistic! When he read the paper to the Royal Society in November 1921, his abstract, as printed in *Nature* (November 24, 1921) emphatically stated, "Statistics obtained by the method of maximum likelihood are always sufficient statistics." And from this it would follow, with the minor quibble that perhaps consistency and asymptotic normality may be needed, that maximum likelihood estimates are always optimum. A truly beautiful theory was born, after over a century and a half in gestation.

Even as the paper was being readied for press, doubts occurred to the one person best equipped to understand the theory, Fisher himself. The bold claim of the abstract does not appear in the published version; neither does its denial. He expressed himself in this way:

> "For the solution of problems of estimation we require a method which for each particular problem will lead us automatically to the statistic by which the criterion of sufficiency is satisfied. Such a method is, I believe, provided by the Method of Maximum Likelihood, although I am not satisfied as to the mathematical rigour of any proof which I can put forward to that effect. Readers of the ensuing pages are invited to form their own opinion as to the possibility of the method of maximum likelihood leading in any case to an insufficient statistic. For my own part I should gladly have withheld publication until a rigorously complete proof could be formulated; but the number and variety of new results which the method discloses press for publication, and at the same time I am not insensible of the advantage which accrues to Applied Mathematics from the co-operation of the Pure Mathematician, and this co-operation is not infrequently called forth by the very imperfections of writers on Applied Mathematics" (Fisher, 1922, page 323).

The 1922 paper did present several related arguments in addition to the Waldian one I reported above. It stated less boldly a converse of the statement in the 1921 abstract that, "it appears that any statistic which fulfils the condition of sufficiency must be a solution obtained by the method of the optimum [e.g. maximum likelihood]" (page 331). But Fisher did not now claim that a sufficient statistic need always exist. Instead Fisher gave an improved non-Bayesian version of the Pearson–Filon argument for asymptotic normality, expanding the likelihood function about the true value and pointing out how and why the argument requires maximum likelihood estimates (and that it would not apply to moment estimates), and how it could be used to assess the accuracy of maximum likelihood estimates (pages 328–329). And there, in a long footnote, he called Karl Pearson to task for not earlier calling attention himself to the error in the 1898 paper. Fisher noted that in 1903 Pearson had published correct standard errors for moment estimates, even while citing the 1898 paper without noting that the standard errors given in 1898 for several examples were wrong. In the 1922 paper Fisher also pointedly included a section illustrating the use of maximum likelihood for Pearson's Type-III distributions (gamma distributions), contrasting his results with the erroneous ones Pearson and Filon had given in 1898 for the same family.

## 6. THREE YEARS LATER

By 1925 Fisher's earlier optimism had faded somewhat, and he prepared a revised version of his theory for presentation to the Cambridge Philosophical Society. At some point in the interim he had recognized that sufficient statistics of the same dimension as the parameter did not always exist. What led to this realization? Fisher did not say, although in a 1935 discussion he wrote, "I ought to mention that the theorem that if a sufficient statistic exists, then it is given by the method of maximum likelihood was proved in my paper of [1922]. . . . It was this that led me to attach especial importance to this method. I did not at that time, however, appreciate the cases in which there is no sufficient statistic, or realize that other properties of the likelihood function, in addition to the position of its maximum, could supply what was lacking" (Fisher, 1935, page 82). I speculate that he learned this in considering a problem where no sufficient statistic exists, namely the problem that figured prominently in the 1925 paper, the estimation of a location parameter for a Cauchy distribution. In any event, in that 1925 paper Fisher did not dwell on this discovery of insufficiency; quite the contrary. The possibility that sufficient statistics need not exist was only casually noted as a fact 14 pages into the paper, and a reader of both the 1922 and 1925 papers might not even notice the subtle shift in emphasis that had taken place.

Where in 1922 Fisher started with consistency and sufficiency, in 1925 he began with *efficiency*. Writing of consistent and asymptotically normal estimates, he stated, "The criterion of efficiency requires that the fixed value to which the variance of a statistic (of the class of which we are speaking) multiplied by $n$, tends, shall be as small as possible. An efficient statistic is one for which this criterion is satisfied" (page 703). With this in mind, his main claim now was (page 707), "We shall see that the method of maximum likelihood will always provide a statistic which, if normally distributed in large samples with variance falling off inversely to the sample number, will be an efficient statistic."

Thus in 1925 the theory said that *if* there is an efficient statistic, then the maximum likelihood estimate is efficient. When a sufficient and consistent estimate exists, it will also be maximum likelihood, but that is not necessary for efficiency. He granted that more than one efficient estimate could exist, but he repeated a proof he had already given in 1924 (Fisher, 1924a) that any two efficient estimates are correlated with correlation that approaches 1.0 as $n$ increases.

## 7. THE 1925 "ANOVA" PROOF

What did Fisher offer by way of proof of this new efficiency-based formulation? His 1922 treatment had leaned crucially on sufficiency, but that was no longer generally available. In its place he depended upon a new and limited but mathematically rather clever proof that I will call the "analysis of variance proof." The proof was clearly based upon a probabilistic version of the analysis of variance breakdown of a sum of squares that Fisher was developing separately at about the same time for agricultural field trials. Fisher's own 1925 presentation of the argument is fairly opaque and does not explain clearly its underlying logic; in 1935 he gave an improved presentation that helps some (Fisher, 1935, pages 42–44). The mathematical details of the proof have been clearly re-presented by Hinkley (1980) at some length. I will be content to offer only a sketch emphasizing the essence of the argument, what I believe to be the logical development Fisher had in mind. It will help the historical discussion to divide his 1925 argument into two parts, just as Fisher did in the 1935 version.

Let $f(x; \theta)$ be the density of a single observation, and let $\phi$ be the likelihood function for a sample of $n$ independent observations, so that $\log \phi = \Sigma \log f$. Following Fisher, let $X = \frac{1}{\phi}\frac{\partial \phi}{\partial \theta} = \frac{\partial}{\partial \theta}\log \phi$—what we now sometimes refer to as the score function. Fisher was only concerned here with situations where the maximum likelihood estimate could be found from solving the equation $X = 0$ for $\theta$. The first part of the argument was really more of a restatement of what he had shown in 1922: from expanding the score function in a Taylor series, he had that the score function was approximately a linear function of the maximum likelihood estimate; as he put it, $X = -nA(\theta - \hat{\theta})$ "if $\theta - \hat{\theta}$ is a small quantity of order $n^{-1/2}$," where his $-nA$ denoted what we now call the Fisher Information in a sample, $I(\theta)$. Since under fairly general regularity conditions $E(X) = \int \frac{1}{\phi}\frac{\partial \phi}{\partial \theta}\phi = \int \frac{\partial \phi}{\partial \theta} = \frac{\partial}{\partial \theta}\int \phi = \frac{\partial}{\partial \theta}1 = 0$, we also have $\mathrm{Var}(X) = I(\theta)$. As Fisher noted, $I(\theta)$ may be found from any of the alternative expressions

$$I(\theta) = -E\left(\frac{\partial^2 \log \phi}{\partial \theta^2}\right) = E\left(\frac{\partial \log \phi}{\partial \theta}\right)^2$$
$$= -nE\left(\frac{\partial^2 \log f}{\partial \theta^2}\right) = nE\left(\frac{\partial \log f}{\partial \theta}\right)^2.$$

Fisher did not discuss conditions under which the linear approximation would prove adequate; he was content to exploit it as a simple route to the asymptotic distribution of the maximum likelihood estimate, namely $N(\theta, 1/I(\theta))$. Thus far he had not gone beyond the 1922 argument.

The part of the argument that was novel in 1925, the "ANOVA proof," then went as follows: Let $T$ be any estimate of $\theta$, assumed to be consistent and asymptotically normal $N(\theta, V)$. In the proof Fisher used this as the exact distribution of $T$, and further treated $V$ as not depending upon $\theta$, as would approximately be the case for "reasonable" estimates $T$ in what we now call "regular" parametric problems. Fisher considered the score function $X$ as a function of the sample and looked at its variation over different samples in two ways. The first was to consider the total variation of $X$ over all samples, namely its variance $\mathrm{Var}(X) = I(\theta)$. And for the second, he evaluated $\mathrm{Var}(X|T)$, the conditional variation in $X$ given the value of $T$ for the sample (i.e., the variance of $X$ among all samples that give the same value for $T$). From this he computed $E[\mathrm{Var}(X|T)]$, which he found equal to $\mathrm{Var}(X) - 1/V$. Since $\mathrm{Var}(X) = E[\mathrm{Var}(X|T)] + \mathrm{Var}[E(X|T)]$ (this is the ANOVA-like breakdown I refer to), this would give $\mathrm{Var}[E(X|T)] = 1/V$. But $\mathrm{Var}(X|T) \geq 0$ always, which implies that necessarily $E[\mathrm{Var}(X|T)] \geq 0$, and so $\mathrm{Var}(X) - 1/V \geq 0$. This gave $\frac{1}{V} \leq I(\theta)$, or $V \geq \frac{1}{I(\theta)}$ for any such $T$, with equality for efficient estimates—what we now refer to as the information inequality. Thus if the maximum likelihood estimate indeed has asymptotic variance $1/I(\theta)$, he had established efficiency.

The logic of the proof—and the likely route that led Fisher to it—seems clear. If there were a sufficient statistic $S$, then the factorization theorem (which Fisher had recognized in 1922, at least in part) would give $\phi = C \cdot h(S; \theta)$, where the proportionality factor $C$ may depend upon the sample but not on $\theta$. By sufficiency, $X$ would then depend upon the sample only through $S$, and so $\mathrm{Var}(X|S) = 0$ for all values of $S$, and consequently $E[\mathrm{Var}(X|S)] = 0$ also. Also, if $S$ is sufficient, the maximum likelihood estimate (found through solving $X = 0$ for $\theta$) is a function of $S$. The failure of $T$ to capture all of the information in the sample is then reflected through the variation in the values of $X$ given $T$, namely through $\mathrm{Var}(X|T)$ and thus $E[\mathrm{Var}(X|T)]$. This latter quantity plays the role of a residual sum of squares and measures the loss of efficiency of $T$ over $S$ (or at least over what would have been achievable had there been a sufficient statistic).

What is more, this interpretation gave Fisher a target to pursue in trying to measure the amount of lost information, or even to determine how one might recover it, just as in an analysis of variance one can advance the analysis by introducing factors that lead to a decrease in the residual sum of squares. In the remainder of the 1925 paper Fisher pursued just such courses. He introduced both the term and the concept of an ancillary statistic, in effect as a covariate designed to reduce the residual sum of squares toward its theoretical minimum achievable value. He gave particular attention to multinomial problems and focused on a study of the information loss when no sufficient estimate existed, and the loss in information in using an estimate that was efficient but not maximum likelihood (e.g., a minimum chi-square estimate). He found the latter difference tended to a finite limit, a measure of what C. R. Rao ([1961], [1962]) was later to term "second-order efficiency."

By 1935 Fisher evidently had come to see the first part of the argument—the part establishing that the maximum likelihood estimate actually achieved the lower bound $1/I(\theta)$—as unsatisfactory, and he offered in its place a different argument to show the bound was achieved. That argument (Fisher, [1935], pages 45–46) was derived from what I will call his third proof; I shall comment on it later in that connection.

Fisher's 1925 work was conceptually deep and has been the subject of much fruitful modern discussion, particularly by Efron ([1975], [1978], [1982], [1998]), Efron and Hinkley ([1978]) and Hinkley ([1980]).

## 8. AFTER 1925: CORRESPONDENCE WITH HOTELLING

Fisher's beautiful theory had become more complicated but was still quite attractive. The proofs Fisher offered in 1925 were not such as would satisfy the Pure Mathematician he had referred to in 1922, nor would they withstand the challenges that would come a quarter century later. Were they all that he could offer? To answer this, it would help us to listen in on a dialogue between Fisher and a non-hostile, highly intelligent party. Many in the audience in England who were interested in this question had axes to wield, and Fisher's transparent digs at Karl Pearson, even though they came in the form of legitimately pointing out major errors in Pearson's previous work, just set those axes a-grinding. But there was one reader who approached Fisher's level as a mathematician and was so distant both geographically (he was in California) and scientifically (he was working on crop estimating at that time) that he was able to engage in just such a dialogue. I refer to Harold Hotelling.

Hotelling received his Ph.D. from Princeton University in 1924, for a dissertation in point set topology. In that same year he joined the Food Research Institute at Stanford University, where he worked on agricultural problems. Soon after, he discovered

Fisher through Fisher's 1925 book, *Statistical Methods for Research Workers*. Hotelling reviewed that book for JASA; in fact he reviewed each of the first seven editions and the first three of these were volunteered reviews, not requested by the Editor (Hotelling, 1951). He started up a correspondence with Fisher, and tried unsuccessfully to get Fisher to visit Stanford in 1928 and 1929 (Stigler, 1999a). After several friendly exchanges of letters, on October 15, 1928, Fisher (who had had several requests from others for detailed mathematical proofs) wrote, asking Hotelling, "Now I want your considered opinion as to the utility of collecting such scraps of theory as are needed to prove just what is wanted for my practical methods." Hotelling replied December 8, strongly encouraging such a work as valuable for mathematics generally, and stated that "a knowledge of the grounds for belief in a theory helps to dispel the absurd notions which tend to cluster even about sound doctrines." Fisher's Christmas Eve 1928 reply proposed that they collaborate:

24 Dec '28
Dear Prof. Hotelling
Your letter has arrived on Christmas Eve, and has given me plenty to think about for the holidays. You will not expect too much of my answer, as you see that I am writing first and thinking afterwards; but I can see already that I have a great deal to thank you for.

After a few hours consideration I believe my right course is to send you a draft contents, to be pulled to pieces or recast as much as you like, and to say I will do my best to fill the bill if you will be joint author and be responsible for the pure mathematics. If you consent to this *and* to taking the first decision, like an editor, as to inclusion or exclusion, on the clear understanding that either of us may throw it up as soon as we think it is not worth while, I will start sending stuff in. It will be mostly new as many of the proofs can be done much better than in my old publications.

Have you all my old stuff? I believe you have, but if not I will try to find anything still lacking.

It seems a monstrous lot of work, but I will not grumble if I need not think too much about arrangement.

Yours sincerely
R. A. Fisher
[Hotelling Papers Box 3]

Fisher's draft table of contents is given as Appendix 1 below. That work was never to be completed. There was no apparent split between the two, but as the project went on, Fisher's increasing focus on genetics as his 1930 book *The Genetical Theory of Natural Selection* went through the press, and Hotelling's move in 1931 to the Department of Economics at Columbia University, were likely causes for the drop in interest. By February 1930 Fisher was writing, "It is a grind getting anything serious done in the way of a text book; I hope you will stick to yours, though; as well as developing the purely mathematical developments." Nonetheless, Hotelling spent nearly six months at Rothamsted over the last half of 1929 and saw quite a bit of Fisher over that time. Hotelling returned to the United States in late December in time to submit a paper to the American Mathematical Society (AMS) and present it at their meeting in Des Moines, December 31. That paper was entitled, "The consistency and ultimate distribution of optimum statistics"; that is, on the consistency and asymptotic normality of maximum likelihood estimates. It was published in the October 1930 issue of the *Transactions* of the AMS.

It is a reasonable guess that the approach taken in the paper reflected Fisher's views to some degree, coming directly after the long visit with Fisher, although Fisher apparently played no direct role in the writing. At any rate, when Fisher wrote to Hotelling on the 7th of January in 1930 to thank him for a copy, Fisher's only complaint was that the definition of "consistency" Hotelling gave was slightly different from Fisher's. Fisher wrote,

"It is worth noting to avoid future confusion that you are using consistency in a somewhat different sense from mine. To me a statistic is inconsistent if it tends to the wrong limit as the sample is increased indefinitely. I do not think I have ever attempted to apply the distinction of consistency or inconsistency to statistics which tend to no limit, whereas you call them all inconsistent. Thus I should not call the mean of a sample from

$$\frac{1}{\pi}\frac{dx}{1 + (x - m)^2}$$

an inconsistent statistic, though you would. Congratulations on a very fine paper."

Hotelling's paper is little referred to today, which seems a shame. It is beautifully written, as was most of Hotelling's work, and among other things he explained Fisher's own work on this topic more clearly than Fisher ever did. He reviewed Fisher's proof of asymptotic normality (the one based upon the Pearson–Filon approach), and he gently noted that "it is not clear what conditions, particularly of continuity, are necessary in order that the proofs which have been given shall be valid." To repair this omission Hotelling offered two explicit proofs for the case of one continuous variable, stating overconfidently that "the extensions to any number of variables are perfectly obvious; and the corresponding theorems for discrete variables follow immediately...." The problem is, as Hotelling's clear exposition makes apparent to a modern reader, the proof does not work. He simplified the problem by transforming the parameter space to a finite interval (if necessary) by an arc tangent transformation, and discretized the observed variable by grouping in a finite number of small intervals, and did not realize that the two combined do not ensure the uniformity he would need to achieve the desired result for other than discrete distributions with bounded parameter sets. The error evidently came to Hotelling's attention by 5 December 1931, when he circulated a list of 37 "Outstanding Problems in the Theory of Statistics." Problem #16 on the list was, "Prove the validity of the double limiting process used in the proof of (Hotelling, 1930), for as general a situation as possible."

## 9. THE GEOMETRIC SHADOW OF A NASTY LITTLE FACT

To this point there had been not even a hint of the future appearance of any nasty, ugly little fact that might sully the beautiful theory. But then, on November 15, 1930, Hotelling wrote to Fisher with some pointed questions. The letter reflected a geometric view of the inference problem that Hotelling seems to have found in Fisher's work by 1926 and developed further after their conversations at Rothamsted. Hotelling gave one statement of the view in his 1930 paper (which must have been drafted at Rothamsted), and he restated it in his November letter in different but equivalent notation. The essence is captured by Figure 2, drawn to display what Hotelling conveyed in words and symbols.

Hotelling considered a parameterized multinomial problem with $m$ cells, where the observations are a vector of relative frequencies of counts $x = (x_1, \ldots, x_m)$ taking values in the $m$-dimensional simplex $\sum_{t=1}^{m} x_t = 1$, $x_t \geq 0$ all $t$. Let the probabilities of the cells $f(p) = (f_1(p), \ldots, f_m(p))$ depend upon a parameter $p$; this describes a curve in the simplex as $p$ varies. Let $p = p_0$ denote the true value of the parameter, let $\hat{p}$ be the maximum likelihood estimate of $p$, and let $f(p_0)$ and $f(\hat{p})$ be the points on the curve corresponding to these two values. In his 1930 paper, Hotelling stated further, "The likelihood $L$ is constant over a system of approximately spherical hypersurfaces about $[x]$. The point $[f(\hat{p})]$ is the point of the curve which lies on the smallest of the approximate spheres meeting the curve, and is therefore approximately the nearest point on the curve to $[x]$" (Hotelling, 1930).

Here then is how Hotelling raised his question in correspondence, in the context of what must have been a shared frame of discourse they had adopted at Rothamsted.

Dear Dr. Fisher:

Thank you very much for your recent letter, with graph and data.

I have been examining various problems in Maximum Likelihood of late; I wonder if you can enlighten me as to the conditions under which your proof holds good regarding the minimum variance of statistics obtained by this method, or rather, as to the exact meaning of the theorem. One of several questions is whether the variance of a statistic or its mean square deviation from the true value should be used as a measure of accuracy.

Denoting by $\hat{p}$ the optimum estimate of a parameter $p$, whose true value is $p_0$, can it be said that the variance of $\hat{p}$, assuming $\hat{p}$ normally distributed, is less than that of any other function of the same observations? Obviously not without further qualification, since a function of the observations can be defined having an arbitrarily small variance. We must therefore restrict the comparison to a special class of functions suitable for estimating $p$, but the definition of this class must not involve $p_0$. How should the class be defined? As the class of consistent statistics? If so, the following difficulty must be faced.

Consider a distribution of frequency among a finite number $m$ of classes, involving a parameter $p$. In a sample of $n$, let $x_t$ be the number [Hotelling evidently means relative frequency] falling in the $t$th class. Let $f_t(p)$ be the probability of an individual falling into this class. If we take $x_1, \ldots, x_m$ as coordinates in $m$-space, the equations

$$x_t = f_t(p) \quad (t = 1, \ldots, m)$$

represent a curve with $p$ as parameter. The points corresponding to samples will form a "globular cluster" (as you so well put it in 1915)[1] about that point on the curve for which $p = p_0$. The method of maximum likelihood corresponds approximately, for large samples, to taking for $\hat{p}$ the parameter of the point of the curve *nearest* to that representing the sample; i.e., to projecting orthogonally. Now consider some other method of projecting sample points upon the curve; for example an orthogonal projection followed by an alternate stretching and contracting along the curve. Then if $p_0$ happens to give a point in one of the regions of condensation [i.e. high density], this method of estimation will, for sufficiently large samples, yield a statistic with smaller variance than that by the method of maximum likelihood. To be sure, its variance will be larger if the true value $p_0$ lies in a region of rarefaction [i.e. low density], and averaging for different possible values of $p_0$ might indicate a greater average variance than that of the optimum statistic. But such an averaging would seem to be of a piece with "Bayes' Theorem," in supposing equal a priori probabilities.

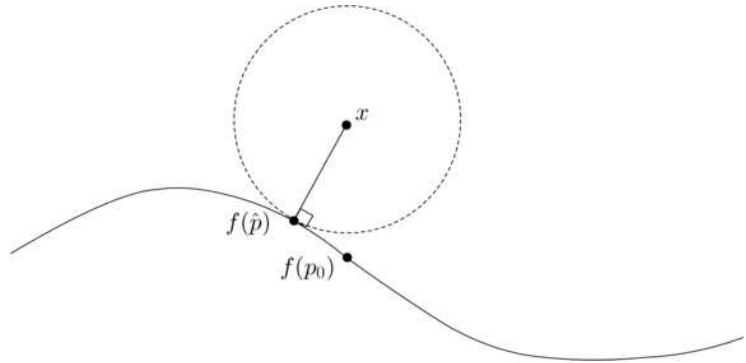

FIG. 2.  *A reconstruction of Hotelling's geometric view of the multinomial estimation problem, circa Fall 1929. Here $x$ represents a multinomial observed relative frequency vector in the simplex, and the curve $f(p)$ the potential values of the multinomial probability vector; the true value of the parameter $p$ ($p_0$) is shown, as is the MLE and a contour of the likelihood surface.*

Hotelling went on to state that even in the particular case of symmetric beta densities, maximum likelihood failed to be optimum, but his derivation there was marred by a simple error in differentiation. Before he received Fisher's reply of the 28th of November, 1930, Hotelling wrote again, on December 12, correcting his own error with regard to the beta estimation problem and enlarging on his other comment, to the point of rather clearly speculating on the possibility of superefficient estimates.

The general question of the exact circumstances in which optimum statistics have minimum variance... is extremely interesting. That the property is not perfectly general seems clear from a consideration of some of the distributions having discontinuities; and also from the fact that, if the true value were known, a system of estimation could be devised which would give it with arbitrarily small variance; and such a system of estimation might happen to be adopted even if the true value were unknown.

I have two students working on the optimum estimates of $m$ for the above curve and for the Type III case you treated. Failing to get anything of consequence for small samples by purely mathematical meth-

---

ods, they will probably soon resort to experiment.[2]

Cordially yours,

Harold Hotelling

Hotelling's letters posed a challenging question in a direct but nonconfrontational way. Clearly, Hotelling said, some more constraints on the class of estimates would be needed; the geometric view they had evidently shared at Rothamsted suggested that consistency alone was not enough. There is no obvious guarantee that the curve $f(p)$ and the contours of the likelihood are such that improvement over maximum likelihood is not possible. What would be needed to prevent this, or at least to convince a reader such as Hotelling that the worry was groundless? Hotelling's hypothetical improvements were certainly vague. A modern reader might be tempted to see them as foreshadowing Hodges's estimate or even shrinkage via Stein estimation, but even though they fall short of that, they presented a clear challenge to Fisher.

## 10. FISHER'S REPLY: A THIRD PROOF OF THE EFFICIENCY OF MAXIMUM LIKELIHOOD

By 1930 Fisher was no stranger to challenges by skeptical readers. His general reaction to one from a friendly source was to state clearly what he was prepared to say, while avoiding speaking directly to the point raised. Without addressing the criticism, much less admitting its validity, he would move directly to a new and improved position, often not giving any indication that it represented the strongest statement that could be made and perhaps even hinting otherwise or at least allowing the reader to speculate so. Such was the case here.

Fisher's reply to the first of Hotelling's letters was brief, but it included one enclosure (A) that outlined a new proof that illuminated Fisher's views, as well as a second short note (B) correcting Hotelling's error in differentiating the beta density.

28 November 1930
Dear Hotelling,

I enclose two notes A and B on the points you raise. The first brings in the general variances and covariances for the multinomial, and is done more prettily by replacing the multinomial by a multiple Poisson,[3] but the argument is probably clearer as it stands.

This is a very short note; the meat of my letter is in the enclosures. . . . .

Yours sincerely,
R. A. Fisher

[Hotelling Papers Box 45]

Fisher's Enclosure A presented a sketch of a new, third proof of the efficiency of maximum likelihood, one that took a different point of attack. The argument, given *in toto* as Appendix 2 below, was elegant, geometric, and I believe also correct, or at least completable, under the tacit regularity conditions implied by the analysis. The geometric stance he took was that which he and Hotelling would have discussed at Rothamsted, restricted to the case of a parametric multinomial family of distributions. Without any discussion, but in evident reply to Hotelling's request for further restrictions on the class of estimates allowed in order to rule out nasty little facts of the sort Hotelling had hinted at, Fisher did introduce a new restriction on the class of estimates $T$ for which the result was claimed. As Fisher put it, the estimates under consideration were now assumed to be homogeneous functions $T = \phi(x_1, \ldots, x_s)$ of degree zero, where $(x_1, \ldots, x_s)$ is the vector of counts. This, and the tacitly assumed smooth differentiability, gave him access to a number of simple relationships leading to a conclusion he summarized as follows:

> "The criterion of consistency thus fixes the value of $T$ at all points on the expectation line, while the criterion of efficiency in conjunction with it fixes the direction in which the equistatistical surface cuts that line. All statistics which are both consistent and efficient thus have surfaces which touch on that line. The surface for Maximum Likelihood has the plane surface of this type."

A homogeneous function $\phi$ of degree $h$ is one where $\phi(cx, cy, \ldots) = c^h \phi(x, y, \ldots)$, and in the applied mathematics of Fisher's day their principal advantage was

---

that if differentiable, they satisfied *Euler's Relation* $x\phi_x + y\phi_y + z\phi_z + \cdots = h\phi(x, y, z, \ldots)$, where $\phi_x$ denotes the partial derivative of $\phi$ with respect to $x$ (see, e.g., Courant, 1936, Vol. 2, pages 108–109). In Fisher's case, the homogeneous functions $\phi$ of degree zero would be functions of the sample relative frequencies only, and would not otherwise depend upon the sample size $n$. This might be considered a strong restriction upon the class of estimates (Fisher did not comment upon this), but with the assumed differentiability and Euler's Relation with $h = 0$, and the exactly known covariances for the multinomial, Fisher had an easy expression for the asymptotic variance for all estimates $T$ in this class. He did not require recourse to the regularity assumptions implicit in the substituting normal distributions for approximately normal distributions, or in assuming the variance of $T$ was approximately constant, as he had in his previous proof. It was then an easy step to use standard Lagrangian methods to minimize this asymptotic expression for the variance for consistent estimates within this class and show the resulting equations were those that also determined the maximum likelihood estimate.

Fisher published this third proof later only in a disguised form, namely where he assumed that the estimate $T$ was to be found from an estimating equation restricted to be a linear function of the relative frequencies; that is, without stating where he had begun, he jumped directly to Euler's Relation. In that guise, and without the geometric setting and intuition, it appeared in his 1935 paper (Fisher, 1935, pages 45–46) where it served to provide an improved version of the demonstration that the maximum likelihood estimate achieves the information lower bound. It also appeared in Fisher (1938, pages 30–32), a 45-page tract he put together for a visit to India at Mahalonobis's invitation in January 1938. That tract was mostly cobbled together from Fisher's papers, and it summarized his view to that time. And in his 1956 book, he gave (apparently only as an illustration) another, simplified version, restricted to estimates $T$ that were themselves linear in the relative frequencies (Fisher, 1956, pages 145–148).

Hotelling's role in this was that of an important catalyst. He helped lead Fisher to reconsider the problem and provided a remarkably acute audience, but he himself did not contribute further to the theory of maximum likelihood. Hotelling did write one other related paper during his nearly six months at Rothamsted in 1929. It was an investigation of the differential geometry of parameter spaces, with what is sometimes called the Jeffreys information metric, after Jeffreys (1946) (Kass, 1989; Kass and Vos, 1997). The paper, entitled "Spaces of statistical parameters," included Type-III or gamma densities as one example, must also have been written by Hotelling at Rothamsted. On 27 December 1929, a summary of the paper was read by Oystein Ore in Hotelling's absence to the Annual Meeting of the American Mathematical Society in Bethlehem, Pennsylvania. Only an abstract was ever published, but the summary Ore read survives (part of a thick folder of other, later notes by Hotelling), and it is printed here, together with the abstract (Hotelling, 1930a), as Appendix 3.

## 11. THE SITUATION TO 1950

In all, Fisher gave three proofs of the optimality of maximum likelihood. The first, in 1922, was based upon the erroneous belief that maximum likelihood estimates were always sufficient statistics, and it depended upon treating approximately normally distributed random variables as if they were in fact normally distributed. The second proof, in 1925, was what I called the ANOVA proof. It too required the same implicit appeal to regularity by using normality in place of approximate normality, as well as assuming that the asymptotic variances of the estimates were approximately constant, and that the likelihood was sufficiently regular to permit the evaluation and manipulation of various integrals. The third, in 1930 in correspondence (and later in print in 1935 in a version restated in terms of estimating functions that lost the geometric origin), placed more severe restrictions upon the distributions (assumed multinomial) and estimates (smoothly differentiable functions of the relative frequencies only, not varying with sample size), but it yielded a more satisfactory proof. Even if not all details were filled in, that task was fairly easy for the limited setting considered. Indeed, the third proof was immune to the ugly little facts that Hotelling hinted at in 1930 and Hodges produced explicitly in 1951, but at a cost in generality. Still, multinomial distributions are as general as one could hope for in the discrete case, and the intuition developed from the geometric setting of that third proof provided at least superficial promise that the result held much more generally, for continuous parametric families.

Over the next few years, several mathematicians recognized the unsatisfactory extent of rigorous support given for such a broad theory and tried their hands at filling in the gaps Fisher had knowingly leapt over as well as some he had not even recognized. The major early efforts were by Joseph Doob (1934, 1936) and Abraham Wald (1943, 1949) in the United States, Daniel Dugué (1937) in France, and then Harald Cramér (1946, 1946a) writing during the war in isolation in Sweden (having read Fisher, Doob, Dugué, but apparently not Wald). Both Doob and Wald had strong connections with Hotelling; both pursued their studies of this topic on Carnegie Fellowships working with Hotelling at Columbia, Doob in 1934–1935, and Wald in 1938–1939. Doob left for the University of Illinois in 1935, but Wald stayed on at Columbia, replacing Hotelling in 1939–1940 while Hotelling was on leave, and again permanently when Hotelling moved to North Carolina in 1946.

Of these writers, Doob and Dugué fell into new difficulties (as Hotelling had in his 1930 paper); Doob was gently corrected by Wald, and Dugué's slip was apparently first noticed a decade later, in the mid-1940s by Edith Mourier, who brought it to Darmois's attention. The Wald and Cramér treatments were the most satisfactory; both raised the level of rigor to new heights, although both suffered from the complexity of the conditions assumed and the limitations imposed. Wald was already publishing on the theory of estimation by 1939, and his 1943 proof of the asymptotic sufficiency of the maximum likelihood estimates can be seen as a form of the completion of Fisher's 1922 proof. Cramér also was firmly based on Fisher; indeed his development followed the structure of Fisher's work closely, but with rigorous demonstrations and explicit statements of conditions. Much of what Cramér presented might be seen as a realization of the book Fisher and Hotelling might have written, albeit without the geometry.

While this reaction to Fisher's theory (namely that it was not true, or at least not proven as stated) progressed in some quarters, another appeared, namely claims that the theory was not new. In this respect the reactions were like those in the seventeenth century to William Harvey's 1628 demonstration of the circulation of blood, where denials of the truth of the claimed phenomenon coexisted with priority claims on behalf of Hippocrates, circa 400 BC (Stigler, 1999, pages 207ff). Karl Pearson to his death and some others in his camp considered Fisher's maximum likelihood simply the Gaussian method, warmed over and served again without overt reference to any Bayesian underpinnings. That can be attributed to a lack of understanding of what Fisher was accomplishing, a phenomenon that afflicted even such first-class older statisticians as G. Udny Yule. Yule's otherwise excellent 1911 textbook was frequently revised but never made more than the most superficial reference to Fisher (other than to his correction to Pearson on degrees of freedom), even to the 10th edition of 1936 (Yule, 1936). Another, more recent claimant's name was added to Gauss's in 1935 when Arthur Bowley, in moving a vote of thanks for Fisher (1935), called attention to work by Edgeworth in 1908–1909 that bore at least superficial similarity to some of Fisher's work, namely the information inequality of the second of the three proofs.

Bowley clearly had only a dim understanding of this work of Fisher's, and his remarks were mild compared to those 15 years later by Jerzy Neyman. Neyman's difficulties with Fisher began in 1934 and involved both scientific and personal issues, in what would become a long-running feud. Generally the dispute simmered at a low level: Fisher would, after the initial split, mostly ignore Neyman except for occasional barbs (usually veiled, without mentioning Neyman by name), and Neyman would generally downplay the importance and originality of Fisher's work, rising on occasion for a more detailed published blast (Zabell, 1992; Kruskal, 1980).

In 1937 Neyman had been content to attribute the simple idea of maximum likelihood to Karl Pearson, citing Pearson's derivation of the product moment estimate of the normal correlation coefficient as the "most probable" value, using the Gaussian method Pearson later abandoned (Neyman, 1937, page 345; 1938, pages 132, 136; Pearson, 1896, pages 262–265). But in 1951 Neyman's focus on Fisher reached a peak, and he latched on to the claim of priority for Edgeworth and deployed it as a rhetorical weapon in the feud. In a review of the collection of papers (Fisher, 1950), Neyman resurrected Bowley's discovery, accusing Fisher of "an unjustified claim of priority" with respect to "the so-called property of efficiency of the maximum likelihood estimates" (Neyman, 1951). What is more, Neyman wrote, "Actually, the proofs of the efficiency of maximum likelihood estimates offered both by Edgeworth and by Fisher are inaccurate, and the assertion, taken at its full generality, is false." This comes close to being an accusation of a false claim of priority for a false discovery of an untrue fact, which would be a rare triple-negative in the history of intellectual property disputes. Savage (1976) wrote of Fisher with this review in mind, "nor did he always emerge as the undisputed champion in bad manners." On the other side, in 1938 Fisher had reviewed Neyman's influential *Lectures and Conferences on Mathematical Statistics* (1938), a book which had only a few grudging references to Fisher. Fisher's review consisted of only two sentences, the first innocuous and the second, "There is not enough original material to justify publication as a book, and too much that is really trivial" (Fisher, 1938–1939). In June 1951, Neyman also wrote to the editor of the *Journal of the American Statistical Association*, W. Allen Wallis, unsuccessfully requesting that the *Journal* reprint Edgeworth's 1908–1909 papers (letter in Neyman papers, Bancroft Library).

But what of the basic question, did Edgeworth precede Fisher and did he in any way influence him if he did? My own view, which is in general accord with the conclusions Jimmie Savage (1976, pages 447–448) and particularly John Pratt (1976) came to from a detailed study of both Edgeworth and Fisher, is that while there was indeed merit to Edgeworth's work on this, there was no merit to the 1951 accusation of "an unjustified claim of priority." In the course of a long, obscure and rambling series of papers emphasizing the use of inverse probability in estimation, Edgeworth did include a treatment of what he called "the direct method free from the speculative character which attaches to inverse probability." He made what can in retrospect be best interpreted as a statement that maximizing the likelihood within a very restricted class of estimates (basically $M$-estimates for location parameters) gives the estimate with smallest standard deviation. The proof he offered (suggested by Professor A. E. H. Love, an expert on the calculus of variations) was based explicitly upon Schwarz's inequality, and bore no resemblance to any Fisher gave.

There is no indication that this work of Edgeworth's ever had any influence upon Fisher or any other worker on this topic. And the obscurity of the prose—uncommonly dense, even by Edgeworthian standards—is such that it is hard to believe the result would have been recognized there by any contemporary reader other than Edgeworth himself. Even at a later time, its recognition required a reader with Fisher's work in hand and either extensive experience with Edgeworth or a strong historical or personal motive. Bowley had studied Edgeworth's work and mode of expression thoroughly in preparing an extended commemorative summary in 1928. Neyman had both historical and personal motives, as well as Bowley's 1935 prompt. Even today anyone who tries to learn what Edgeworth accomplished from Bowley's 1928 summary (Bowley, 1928, pages 26–28) would emerge completely at sea, no matter how long the text is puzzled over. This is not to deny that when one has dug through the thicket of the 1908–1909 original, there is a limited result and a hint of understanding that went beyond the limited result. Edgeworth was a statistical scientist with an uncommonly subtle and deep mind (Stigler, 1986, Chapter 9; 1999, Chapter 5), and his work here is further evidence of that. But, for all that, the work stands as an independent partial anticipation—a hint, not an instance, of what was to come.

Edgeworth died in 1926 without ever commenting on Fisher, and Fisher, as was his wont, dug in his heels and refused to seriously engage the issue in print. His frankest private statements were in two letters. The first was a 12 February 1940 letter to Maurice Fréchet, where he described Edgeworth's statement as confusingly linked to inverse probability, even though the mathematics could be dissociated from that approach. In that letter Fisher summed up his view in these words: "The confusion of associating this method with Bayes' theorem seems to have been originally due to Gauss, who certainly recognized its merits as a method of estimation, though I do not know whether he proved anything definite about it. I do not know of any explicit statement of the properties, consistency, efficiency and sufficiency, which may characterize estimates prior to my 1922 paper" (Bennett, 1990, page 125). The second letter, dated 2 July 1951, was to a Californian, Horace Gray, who had spent time with Fisher in London 1935–1936, and he had written to Fisher to call attention to Neyman's review. Fisher replied,

> "Neyman is, judging from my own experience, a malicious mischief-maker. ... Edgeworth's paper of 1908 has, of course, been long familiar to me, and to other English statisticians. No one could now read it without realizing that the author was profoundly confused. I should say, for my own part, that he certainly had an inkling of what I later demonstrated. The view that, in any proper sense, he anticipated me is made difficult by a number of verifiable facts" (Bennett, 1990, pages 138–139).

The facts Fisher listed were that (i) Edgeworth based his investigation on inverse probability, (ii) he limited attention to location parameters, and (iii) the formula they shared in common, for the variance of efficient estimates, had been drawn from Pearson and Filon with no notice given to the major errors in that work. Fisher noted that since by 1903 Sheppard's works had shown that moment estimates had variances different from those given by Pearson and Filon, this to Fisher raised the questions: "Had Pearson and Filon's variances any validity at all? Does any class of estimate actually have these variances? If so, how can such an estimate be obtained in general? But Edgeworth would have been far ahead of his time had he asked them." Fisher would grant Edgeworth "an inkling," but no more. Some might see more in Edgeworth than Fisher did, but they do so from a different historical perspective. I believe Fisher owed no intellectual debt to Edgeworth on this issue, and it was his own loss. Had he taken the time and trouble to learn from Edgeworth's insight, he might have gone even further. Savage (1976) proffered as explanations for this neglect, that Fisher initially thought Edgeworth's premises ridiculous, and later "because it is hard to seek diligently after the unwelcome."

Neyman's was not the only review to raise priority issues about Fisher's work. In a tendentious 1930 review of the 3rd edition of Fisher's *Statistical Methods for Research Workers*, Charles Grove seemed to claim that all in Fisher was to be found earlier in Scandinavian work by Thiele, Gram or Charlier. Grove (1930) did not focus on maximum likelihood, which he evidently thought was unsupported, but put forth instead the claim that Thiele had in 1889 anticipated Fisher on small sample inference and particularly on estimating cumulants with $k$-statistics, and Gram had done so on the use of orthogonal polynomials in regression. Fisher replied in the same publication, and more colorfully in a private letter to Grove's colleague Arne Fisher (a Dane who seems to have been the instigator of Grove's review). Fisher stated that Thiele "had no more glimmer than [Karl] Pearson of some of the ideas we now use" (Grove, 1930; Fisher, 1931; Bennett, 1990,

page 313). A scrupulous recent translation of Thiele from the Danish (Lauritzen, 2002) with accompanying commentary allows a better assessment of his excellent work, which, however, did not include contributions to maximum likelihood estimation.

## 12. DOUBTS ABOUT MAXIMUM LIKELIHOOD

The possibility that maximum likelihood estimates could actually perform badly, or that they might be dramatically improved upon by another method, seems to have not been raised prior to Hotelling's probing letters to Fisher of November 15 and December 12, 1930. Kirstine Smith and Karl Pearson had questioned the relative merits of the "Gaussian method" versus minimum chi-square in 1916, but any difference there was minor; both were later seen to be asymptotically efficient estimates. For the most part, the early reservations about Fisher's maximum likelihood centered on questions of priority (was he preceded? was anything really new in the method?) and issues of practical usefulness (were the calculations too hard relative to the method of moments?). As maximum likelihood became more widely adopted in the 1930s, the increased attention to proofs of its effectiveness (could a rigorous general demonstration be devised?) led inevitably to questions of when it might break down. The earliest explicit example is perhaps due to Abraham Wald, in correspondence with Jerzy Neyman in 1938.

Wald immigrated to the United States from Vienna in Spring 1938, when shortly after Hitler's annexation of Austria he accepted an offer to join the Cowles Commission for Research in Economics, then located in Colorado Springs. He remained with Cowles through the summer before joining Harold Hotelling at Columbia University in Fall 1938. On September 20, 1938, a week before he left for Columbia, Wald wrote to Neyman sending a promised manuscript on the Markov inequality, but also describing a different problem he had encountered. The problem described was a slight generalization of one he would treat in Wald (1940), namely estimating a straight line when $n$ points on the line are observed but both coordinates are subject to independent errors. However, Wald's letter to Neyman contained a statement that he omitted from the 1940 article: "I have shown that the method of maximum likelihood leads to *false estimations* of the parameters.... (i.e., leads to statistics of which the stochastic limits are unequal to the values of the respective parameters to be estimated). Hence the maximum likelihood method cannot be applied" (Neyman Papers, Box 14, Folder 28). Wald stated that he had solved this general estimation problem for the case of independent normally distributed errors with possibly unequal variances.[4]

Neyman replied on September 23 that he was quite interested in the new problem, "the more so as it is rather close to what I am trying to do myself." Ten years later Neyman and Elizabeth Scott published, with a general citation to Wald (1940), a simplified version of Wald's example as one of several with increasing numbers of parameters where maximum likelihood estimates are inconsistent. That version, in which the straight line $y = x$ and the two coordinates' error variances are equal, has come to be known as the Neyman–Scott example. It is usually expressed as follows: $X_{ij}$ are independent $N(\mu_j, \sigma^2)$, for $i = 1, 2$, and $j = 1, \ldots, n$, in which case the maximum likelihood estimate of $\sigma^2$ consistently estimates half the correct value (Neyman and Scott, 1948).

In June of 1951, just as Jerzy Neyman's review of Fisher's *Collected Papers* appeared, the Berkeley Statistical Laboratory convened for the summer under Neyman's general direction. One of three research groups took as its charge "a complex of questions arising from considerations of superefficiency and identifiability." The group concentrating on this topic was comprised of Joseph L. Hodges, Jr., Lucien Le Cam and Agnes Berger. It was presumably shortly before that time that Hodges, then an Assistant Professor at Berkeley, constructed his example; in any event the study was soon sufficiently advanced that a session on the topic "Efficiency and superefficiency of estimates" was arranged by Neyman to be held on Saturday, December 29, 1951, at the Boston meeting of the Institute of Mathematical Statistics. Four talks were presented in that session, by Jerzy Neyman ("On the problem of asymptotic efficiency of estimates"), Joe Hodges ("Local superefficiency"), Lucien Le Cam ("On sets of parameter points where it is possible to achieve superefficiency of estimates") and Joseph Berkson ("Relative precision of least squares and maximum likelihood estimates of regression coefficients") (*Biometrics*, 1951;

Littauer and Mode, 1952). Neither Neyman's nor Hodges's talks were ever published; Le Cam's was developed into his Ph.D. dissertation and published in 1953. That publication (Le Cam, 1953) included Hodges's example (credited to Hodges), and Le Cam proved among other things that while superefficiency was clearly possible, the set of parameter points where it could be achieved had Lebesgue measure zero.

In the decade that followed, a number of other examples were discovered or devised. Of these, the least contrived was the problem of estimation for the five-parameter mixture of two normal distributions, where the likelihood function explodes to infinity when either mean parameter equals any observation. This and several other examples, including an important one by Bahadur, are reviewed in Le Cam (1990) and Cox (2006, Chapter 7). Le Cam speculates that the normal mixture example (known in the folklore of the 1950s but apparently not published then) was due to Jack Kiefer and Jacob Wolfowitz; Cox (2006, pages 134–135) considers it to some extent pathological.

These early examples created a flurry of excitement but are for the most part not seen today as debilitating to the theory. Hodges's example made a substantial impact when it first became known, but it has, ever since Le Cam's dissertation, come to be seen as an ingenious but minor technical achievement. Hodges (see Figure 1 above) showed you could improve locally on maximum likelihood, basically by shrinking the estimate toward zero, and as such it might also be viewed as an early hint of the 1955 shrinkage estimates of Charles Stein that in multiparameter problems can improve uniformly on maximum likelihood. But Hodges's example itself was for finite samples inferior to maximum likelihood for parameter values not near zero, and it was not long seen as a serious threat. The Wald–Neyman–Scott example was of more practical import, and still serves as a warning of what might occur in modern highly parameterized problems, where the information in the data may be spread too thinly to achieve asymptotic consistency. The normal mixture example remains certainly of at least computational importance, as showing how in complex settings it may be necessary to seek local maxima or to constrain the parameter space. Fisher never commented on any of these examples.

There continued over this period to be a number of attempts to complete the theory, to give a

---

rigorous description of conditions that approached necessary and sufficient, conditions describing situations in which maximum likelihood would not mislead. As work on the topic became more refined and more correct, the intrinsic difficulties of the topic also became more apparent. The lists of conditions needed to prove optimality by Wald and Cramér were already unwieldy and the basic logic of the solutions retreated from sight; indeed one problem was that achieving rigor sometimes led to the exclusion of basic examples, such as the estimation of the normal standard deviation, as in Wald (1943). The consequences can still be seen today, in the best textbook treatments, such as those by Bickel and Doksum (2001) and by van der Vaart (1998), where the elegance of the exposition comes from strategically restricting the range of the coverage. Bahadur (1964) gave a succinct and elegant theorem that builds upon work of Le Cam, but only treated a one-dimensional parameter and was restricted to estimates that are asymptotically normal with variances that are continuous in the parameter.

## 13. OF ERRORS IN THEORY

At many junctures in this story we have encountered what might be judged theoretical errors committed by the workers involved. Perhaps Lagrange, by ignoring the curve his probabilities followed until the final stage, could be judged in error; it certainly left him with method of moment estimates that would be thought woefully inefficient by the Fisher generation. Perhaps Gauss's use of a uniform prior, which rendered his solution susceptible to change by nonlinear transformations of the parameters, would be considered an error. Certainly Pearson and Filon erred in their promiscuous use of a naïve passage to a limit in ways where it gave wrong answers (Stigler, 2007). And certainly Fisher's 1921 assumption that sufficient statistics always exist was an error, and Hotelling's 1930 proof of the consistency and asymptotic normality of the maximum likelihood estimate cannot be counted correct for the generality claimed.

There are other errors I have not discussed. When Lambert (1760) in a sketchy presentation gave only one example, he got what was arguably the wrong answer there. Lambert's only specific result was for $n = 2$, claiming the sample mean in that case always gave the most probable result, a claim that would fail for the Cauchy density. See Stigler (1999, Chapter 16). And there were later smaller and subtler lapses in rigor in attempts by Doob and by Dugué to themselves correct some of Fisher's oversights. But I do not mean at all to suggest these pioneers had feet of clay. To the contrary. Without Lagrange's error he might not have found the Laplace transform at that early date. Without Pearson and Filon, Fisher might not have started down the road he did. Without Fisher's 1921 mistaken jump to a conclusion, he might not have rushed to complete his theory, which even flawed and incomplete, was instrumental in launching twentieth century theoretical statistics. Great explorations in uncharted territory seem to require great boldness, and even mischance can lead to major advance.

## 14. CONCLUSION

Despite all these difficulties, maximum likelihood remains one of the most used and useful techniques of modern statistics. How can that be, in the face of the nasty little facts uncovered by the 1950s? For one thing, there is solid mathematical support in a wide class of problems. Fisher's proofs can all be defended as correct, at least if one accepts as given the regularity conditions and assumptions that were clearly implicit, including the limitation in 1922 to sufficient estimates, and in 1925 to score functions linearly approximable by maximum likelihood estimates. Of course that defense flirts with tautology: any statement is true if all the conditions required for its truth are assumed; even the Pearson–Filon derivation of the "probable errors of frequency constants" might be so defended. But there is a big difference between the two cases. Fisher's implicit assumptions are in part fairly clear (smooth differentiability, consistent estimates, e.g.) and were clearly evident to Fisher himself; if he made a false application of the theory, it is not known to me. On the other hand, with Pearson–Filon the case was different, as the inappropriate applications in the same paper make clear. Nonetheless, the intense mathematical investigations after 1938 and particularly in the 1950s revealed potential problems Fisher had not considered, with increasing numbers of parameters, unbounded likelihood functions and the possibility of local improvement over maximum likelihood. Fisher was surely aware of some of these problems, at least when they were published, if not before. The first of them he might have countered by noting that in such situations the amount of information in the data (the measurement of which was one of his pioneering advances) was being spread over a space of dimension increasing in proportion with the sample size, and so of course problems with consistency could be expected. But he did not. The third possibility, local improvement, had been brought early to his attention by Hotelling, but here too Fisher remained silent, as he did in the face of other examples as well. An explanation for this silence might, ironically, have been given by Fisher himself in a 14 January 1933 letter to Egon Pearson, commiserating with him on the difficulties he faced with his father, Karl: "Many original men are for that reason unreceptive, and this is a fault which age does nothing to cure" (Fisher papers).

Personalities played a role in this development. Fisher's hostilities with Neyman surely increased his stubborn resistance to public discussion of areas where questions remained, and they surely contributed to the zeal with which Neyman pursued the discovery and public discussion of such problems. The latter might be viewed as a benefit of the feud: when peace reigned in the early 1930s and the only attention to the problem was by Fisher, Hotelling, and those Hotelling inspired to work on this (Doob, Wald) or a noncombatant (Dugué), the problems in the proofs and the limitations of the theory were not on public view. Indeed, there has been no published criticism that clearly identified the source of errors in the proofs of Hotelling, Doob or Dugué even to the present day; either the early works were ignored, were merely cited, or referred to with a polite allusion such as to the proofs being "not rigorous" (e.g., Doob, 1934; Le Cam, 1953). The reader got no sense of where and how real problems with the theory might arise. Hostility bred uncivil discourse; it also led to principled focus.

Yet despite these problems, time and again maximum likelihood has proved useful even in situations where no general theorem could be found to defend its use. Perhaps as Fisher's powerful geometric intuition may have foreseen, the scope of useful application of maximum likelihood exceeds that of any reasonably achievable proof, even though this comes at the potential cost of inadvertently blundering into a region of inapplicability. We now understand the limitations of maximum likelihood better than Fisher did, but far from well enough to guarantee safety in its application in complex situations where it is most needed. Maximum likelihood remains a truly beautiful theory, even though tragedy may lurk around a corner.

## APPENDIX 1: FISHER'S DECEMBER 1928 DRAFT TABLE OF CONTENTS [HOTELLING PAPERS, BOX 3]

I. Distributions
Varieties and variables
Types of distribution

(a) Discontinuous, step like integrals
(b) Continuous, differentiable integrals
(c) General type, integral not differentiable but frequency not confined to zero measure

Specification by moments
Characteristic function, $\int e^{itx} f(x)\, dx$ or $\int e^{itx}\, dF(x)$
Its logarithm, cumulative property
Cumulative moment functions or seminvariants [sic]
Illustrative cases, uniqueness of normal distribution, multinomial and multiple Poisson

II. Distributions derived from normal
$\chi^2$ distribution is that of $S_1^n(x_p^2)$ when $x_p$ is distributed with unit variance about zero
Transformation of $\xi_q = \sum_{p=1}^n c_{pq} x_p$, $\sum_{p=1}^n c_{pq}^2 = 1$, $\sum_{p=1}^n c_{pq} c_{pq'} = 0$
Application of $\chi^2$ to frequencies
Distribution of $t = \frac{n\bar{x}}{\chi^2}$ [sic]; application to regression coefficients; of $z = \frac{1}{2}\log\frac{n_2\chi_1^2}{n_1\chi_2^2}$.

III. Distribution of correlation coefficient, partial correlation, multiple correlation. Hyperspace treatment

IV. Moment estimates of seminvariants
Simple and multiple distribution of such estimates {paper in Lond. Math. Soc. They will not publish for a year if then}
Combinatorial method

V. Theory of estimation
(Much as already done but more about Sufficient Statistics)
Method of maximum likelihood
Bayes' theorem. Inverse probability and likelihood. Illustrate by inefficiency of moments with Pearsonian curves.

VI. Experimental design (not so agricultural as in Statistical Methods for Research Workers), more use of *amount of information*

VII. Statistical mechanics; argument put in a clear light without taking $x$! as a continuous function when $x$ is small! NOT YET DONE!
Analogous biological problems. Rather worth doing though.

Fowler has now done a great deal, but still the method of steepest descent seems very indirect, and obviously this limits statistical argument.

## APPENDIX 2: FISHER'S ENCLOSURE A FROM THE NOV. 28, 1930 LETTER TO HOTELLING. THE LETTERS WERE TYPED BUT THE FORMULAS WERE WRITTEN IN BY HAND, AND THE APPARENT TYPOGRAPHICAL ERROR IN THE FOURTH FORMULA FROM THE BOTTOM ($\frac{\partial \theta}{\partial X_1}$ FOR $\frac{\partial \phi}{\partial X_1}$) IS AS WRITTEN IN THE ORIGINAL [HOTELLING PAPERS, BOX 45]

Expectation line $x = f(\theta)$

Equistatistical surface (or region) $T = \phi(x_1, \ldots, x_s)$

$$\sum x \frac{\partial \phi}{\partial x} = 0 \qquad \text{if } \phi \text{ is homogeneous of zero degree.}$$

For consistency $\theta = \phi(f_1, \ldots, f_s)$

For large samples, provided there is no bias of order as high as $n^{-1/2}$,

$$V(T) = \text{Mean} \left( \sum \frac{\partial \phi}{\partial x} \delta x \right)^2$$

$$= \sum f \left( 1 - \frac{b}{x} \right) \left( \frac{\partial \phi}{\partial x} \right)^2 - \sum \sum \frac{f f'}{n} \frac{\partial \phi}{\partial x} \frac{\partial \phi}{\partial x'}$$

for multinomial, where differentials refer to the expectation point.

Differentiating the condition of consistency, $d\theta = (\sum \frac{\partial \phi}{\partial x} \frac{\partial f}{\partial \theta}) \, d\theta$, or $\sum \frac{\partial \phi}{\partial x} \frac{\partial f}{\partial \theta} = 1$

Any values $\frac{\partial \phi}{\partial x}$ are admissible subject to this condition for consistency, we may therefore minimize the expression for the variance subject to this condition and obtain equations of the form

$$f_1(\theta) \frac{\partial \theta}{\partial x_1} - \frac{f_1}{n} \sum f \frac{\partial \phi}{\partial x} = \lambda \frac{\partial f_1}{\partial \theta}$$

Now if $\phi$ is homogeneous in $x$ of zero degree, $\sum f \frac{\partial \phi}{\partial x} = 0$, hence, for all classes

$$\frac{\partial \phi}{\partial x} = \frac{\lambda}{f} \frac{\partial f}{\partial \theta} \text{ or}$$

$$\frac{\partial \phi}{\partial x} = \frac{1}{f} \frac{\partial f}{\partial \theta} \Big/ \sum \frac{1}{f} \left( \frac{\partial f}{\partial \theta} \right)^2$$

The criterion of consistency thus fixes the value of $T$ at all points on the expectation line, while the criterion of efficiency in conjunction with it fixes the direction in which the equistatistical surface cuts that line. All statistics which are both consistent and efficient thus have surfaces which touch on that line. The surface for Maximum Likelihood has the plane surface of this type.

## APPENDIX 3: HOTELLING ON PARAMETER SPACES

Hotelling briefly attended the American Mathematical Society's Annual Meeting in Bethlehem, Pennsylvania, December 26–29, 1929, but he left before this paper was scheduled to be read on December 27. The paper was read in his absence by Professor Oystein Ore of Yale University, and Ore subsequently returned the manuscript to Hotelling; only the abstract was ever published (Hotelling, 1930a). Meanwhile, Hotelling traveled on to the AMS Regular Meeting December 30–31 in Des Moines, Iowa, where on December 31 he read his paper "The consistency and ultimate distribution of optimum statistics" (Hotelling, 1930). The summary that follows is the entire manuscript as read by Ore, from the Hotelling Papers at Columbia University (Box 44).

### SPACES OF STATISTICAL PARAMETERS

By Harold Hotelling, Stanford University.
[Abstract]

For a space of $n$ dimensions representing the parameters $p_1, \ldots, p_n$ of a frequency distribution, a statistically significant metric is defined by means of the variances and co-variances of efficient estimates of these parameters. Such a space, for the ordinary types of distributions, is always curved. For the two parameters of the normal law the manifold may be represented in part as a surface of revolution of negative curvature, with a sharp circular edge. On this surface variation of the dispersion is represented by moving along a generator. For a Pearson Type III curve [i.e. gamma distributions] of any given shape the same surface occurs. For the unrestricted Type III curve there are three parameters; their space is investigated. Certain metrical properties which hold in general for spaces of statistical parameters are given.

### SUMMARY OF "SPACES OF STATISTICAL PARAMETERS"

A "population" is specified by a function

$$f(x, p_1, \ldots, p_k)$$

such that $f\,dx$ is the probability of an observation falling in the range $dx$. In statistical theory we have given observations $x_1, \ldots, x_N$ and wish to estimate the values of the parameters $p_1, \ldots, p_k$. There is an infinity of possible methods of making these estimates; but one possessing certain peculiarly valuable properties is that of *maximum likelihood*. The likelihood is defined as

$$\prod_{i=1}^{N} f(x_i, p_1, \ldots, p_k).$$

Denote its logarithm by $L$. Let $\hat{p}_1, \ldots, \hat{p}_k$ be the values maximizing $L$. They have been called *optimum statistics*, or optimum estimates of the parameters, by R. A. Fisher. The errors of estimate $\hat{p}_\alpha - p_\alpha$ derived from samples of $N$ have a distribution which for large values of $N$ approaches the normal form

$$K e^{-\frac{1}{2}T} d\hat{p}_1, \ldots, d\hat{p}_k,$$

where

$$T = \sum \sum g_{\alpha\beta}(\hat{p}_\alpha - p_\alpha)(\hat{p}_\beta - p_\beta).$$

Here $g_{\alpha\beta}$ is the mathematical expectation of

$$\frac{\partial^2 L}{\partial p_\alpha \, \partial p_\beta},$$

and is a covariant tensor of second order under transformations $p'_\alpha = \phi_\alpha(p_1, \ldots, p_k)$—though of course the second derivative is not itself a tensor.

This tensor property suggests that

$$g_{\alpha\beta} \, dp^\alpha \, dp^\beta$$

be taken as distance element in a space of coordinates $p^1, \ldots, p^k$. Indeed a considerable amount of differential geometry carries over immediately to give novel statistical conclusions. It should be said at once that these spaces are not flat, but are curved in a manner depending on the initial population distributions.

Problems of "random migration" by short leaps in the $k$-space occur in various biological problems, when evolution is supposed to take place by small mutations. Such problems occur also in experimental work, as in the dilution method of counting soil bacteria developed by Cutter at Rothamsted. These problems, for short steps, are equivalent to problems regarding heat conduction and geodesics in the curved space.

If we are considering an initial distribution curve of any fixed *shape*, we have two parameters to estimate, giving the *location* and *scale* of the curve, for example the mean and standard deviation of a normal error curve. Our $k$-space is in such cases a surface of constant negative curvature. Representing the normal curve by means of a pseudosphere, variation of the standard deviation is represented by motion along a generator, variations of the mean by rotation about the axis. A greater variance means closer propinquity to the axis.

Since a geodesic on a pseudosphere between two points on the same meridian comes closer to the axis than the meridian, we have an interesting biological conclusion. If we have two related species having about the same variance but a difference in means, the most likely common ancestors had a greater variance than either existing species.

For a Pearson Type III curve the measures of position and scale vary, not along geodesic but along loxodromes.

Spaces of statistical parameters lend themselves to the treatment of a wide range of problems in which discrepancies between hypothesis and observation which involve two or more observations are to be tested. Thus if the hypothesis to be tested is that a species, in which the frequency distribution of some dimension has the normal form, has arisen by a succession of small mutations from another, and if we consider the difference of the variances along with that of the means, we are led to apply the distribution of $\chi^2$ for $n = 2$, just as in judging marksmanship we may combine vertical with horizontal deviations of a shot from the center of the target. But the fact that the surface, on which the mean and variance are coordinates, is a pseudosphere instead of a plane, shows that a correction must be applied to the probability of a greater deviation as calculated from $\chi^2$. Indeed, the area or circumference of a geodesic circle is *greater* than for one of the same radius on a plane. The excess of area measures the correction which must be applied to obtain the true probability of a greater discrepancy.

If about a point on the pseudosphere representing any population we describe a geodesic circle, the points on the circumference represent statistics, such as mean and variance, which might with *equal likelihood* have been obtained in a sample from this population. And inversely, if corresponding to a given sample, we fix upon a point as center of a geodesic circle, the points on the circumference represent populations which, on the evidence of this sample, are all equally likely.

## ACKNOWLEDGMENTS

I am grateful to Henry Bennett for permission to consult and quote from the Fisher Papers in Adelaide, to Michael Ryan for permission to quote from the Harold Hotelling Papers, Rare Book and Manuscript Library, Columbia University, and to Susan Snyder for permission to quote from the Jerzy Neyman Papers (Call Number BANC MSS 84/30C, Box 14, Folder 28), Bancroft Library, University of California Berkeley. For comments during the course of this investigation, I thank Peter Bickel, Larry Brown, Bernard Bru, David Cox, Persi Diaconis, Anthony Edwards, Brad Efron, Tim Gregoire, Marc Hallin, Lucien Le Cam, Erich Lehmann, Peter McCullagh, Edith Mourier, Ingram Olkin. The paper is based upon material presented as the Lucien Le Cam Memorial Lecture at the IMS Annual Meeting in Rio de Janeiro, August 2, 2006.

## REFERENCES

Norden (1972–1973) surveys the literature to 1972, with an extensive bibliography. Hald (1998) gives a detailed report in modern notation of Fisher's published work in statistical inference, as well as work related to maximum likelihood by Pearson and Filon and by Edgeworth. Aldrich (1997) and Edwards (1997a) discuss Fisher's earliest work on maximum likelihood; I also describe this work from a different perspective in Stigler (2005), and other aspects of Fisher's work in Stigler (1973, 2001, 2007). Edwards (1974), Kendall (1961) and Hald (1998, 2007) are among those who describe the predecessors of maximum likelihood; see Stigler (1999, Chapter 16), for more detail on this and further references. Savage (1976), Pratt (1976) and Hald (1998, 2007) give accounts of the relationship of Fisher's work to Edgeworth's. For the best appreciation of the role of geometry in Fisher's work on estimation see Efron's Wald Lecture (1982), and for an elegant and insightful modern development of Fisher's geometric approach with historical references, see Kass and Vos (1997). On Hotelling's life and work see Arrow and Lehmann (2005), Darnell (1988), Hotelling (1990) and Smith (1978); on Fisher's life see Box (1978); on Pearson's life see Porter (2004). The Hotelling–Fisher correspondence is housed in the Hotelling Collection at Columbia University (Rare Book and Manuscript Library) and the Fisher Papers at the University of Adelaide; each has a more complete set of the received letters than the sent letters. Neyman's papers are at the Bancroft Library of the University of California Berkeley.